# Study on the preparation of nanosulfur/bentonite complex

Pak Gyu Hui[1], Choe Yong A[1], Son Yong Chol[2], Pong Chol Ung[2] & Sin Kye Ryong[2]*

[1] Analysis Institute of **Kim Il Sung** University

[2] Faculty of Chemistry, **Kim Il Sung** University

Pyongyang, Democratic People's Republic of Korea

*Email Address: ryongnam9@yahoo.com

**Abstract**

Inorganic compounds with sulfur have a good antifungal efficacy. However there is a limit, sulfur is required in bulk quantities for application. Nanosulfur might have a high anti-bacterial effect in a thin concentration. Nanosulfur-bentonite composite was prepared from nanosulfur made by the reaction between sodium thiosulfate and sulfuric acid and bentonite was used as a carrier. The most appropriate reaction temperature, concentrations of $Na_2S_2O_3\ 5H_2O$ and $H_2SO_4$, dropping rate and stirring one were determined. From SEM images, it was illuminated that small particle size of nanosulfur deposited on bentonite is in the range of 20~30nm and big one is in the range of 60~100nm.

1. Introduction

Inorganic germicides with sulfur, e.g. vulcanization accelerator of sulfur, smoke agent of sulfur, lime-sulfur mixture and hydration agent of sulfur, etc. , have a high anti-bacterial actions against powdery mildew germ and rust fungi which has a lot of lipoprotein in fungus body.

Sulfur is lipophilic materials, thus it dissolve the lipid of mycelium and thus exert an anti-bacterial functions, penetrating into cells.

Especially nano-sulfur have enhanced adhesive nature and penetration abilities, so that it is expected that small amount might have high anti-bacterial effect. But it need a lot of agents and complex process in order to make nano-sulfur and solidization of sulfur. Until now a lot of activities to investigate the new method to make a nano-sulfur with simple and safe process.

In this paper we considered the effects of several factors on the preparation when nanosulfur-bentonite complex is produced as a multi-ability nanosulfur germicides by immersing nanosulfur produced from reaction between sodium thiosulfate and sulfuric acid into bentonite and established the reasonable preparation method.

2. Materials and methods

Here used were a 500mL round-bottom flask equipped with a stirrer, VO 200 vaccum drier, stream of water pump and then BT-9300 Lazer Particle size analyzer and JSM-6610A Scanning electron dispersive spectrometer.

All reagents used were of AR grade.

Sodium thiosulfate($Na_2SO_3$ $5H_2O$) and 30% sulfuric acid solution($H_2SO_4$) were used.

The experiment procedures were performed as follows:

100mL of 0.01mol/L $H_2SO_4$ solution and the proper amount of bentonite were transferred into the flask for reaction.

0.03mol/L $Na_2SO_3$ $5H_2O$ solution was come down in drops at a rate of twenty drops per minute.

It was stirred for 3 hours and then enclosed for 4 hours at room temperature.

The precipitation solution with the white color was filtered by using stream pump and then washed for 3 times with the deionized water until $SO_4^{2-}$ ion was not detected in that solution.(Until the white precipitate was not appeared when 0.02mol/L $BaCl_2$ solution was dropped)

Nanosulphur complex was obtained by using vaccum drier from 70℃ to 80℃ for 2 hours from the washed precipitate.

Particle size of nanosulfur-bentonite complex obtained, the surface state of bentonite and particle size of nanosulfur immersed in bentonite was determined by a BT-9300 Lazer analysis and JSM-6610A/EDS.

3. **Results and discussion**

In this work the effect of several fectors on the preparation of nanosulfur-bentonite complex was examined.

Sulfuric acid solution was added into sodium thiosulfuric acid solution and the chemical reaction performed is following as:

Here sulfur particle size was changed with the concentration of reactants and the reaction temperature.

First, the effect of reaction temperature on the average diameter of nanosulfur was observed under the condition that the concentration of sodium thiosulfuric acid was 0.03mol/L, sulfuric acid was 0.01mol/L, the reaction time was 3 hours, the dropping rate was 50 drops per minute and the stirring rate was 1000rpm, was shown in figure 1.

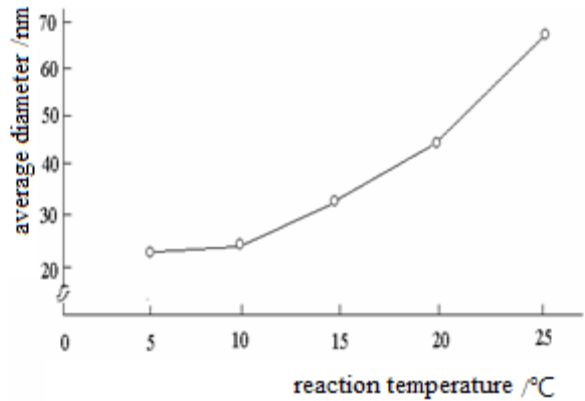

Figure 1. average diameter change of nanosulfur depending on the reaction temperature

As can be seen from figure 1, the higher room temperature was, the greater particle sizes of nanosulfur obtained were.

It is associated with the serious aggregation, it occurred due to the number of collision increased, activating the Brownian motion of particles produced as high as room temperature is.

Thus, the reaction temperature must be as low as possible, but it was better that the reaction was performed at the temperature from 5°C to 10°C in consideration of the reaction rate.

Next, the effect of concentration of concentration of sodium thiosulfate solution on the average diameter of nanosulfur was observed under the some condition by changing concentration of sodium thiosulfate solution at the reaction temperature of 10°C, was shown in figure 2.

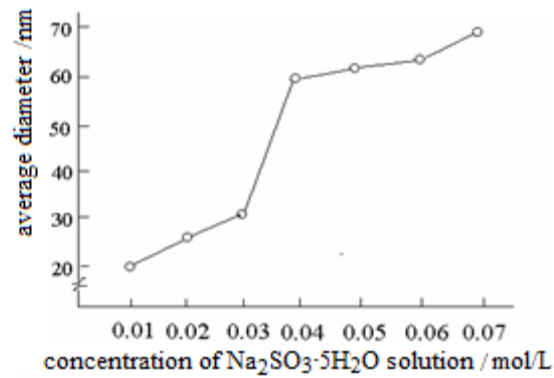

Figure 2. average diameter change of nanosulfur depending on the concentration of $Na_2SO_3 \cdot 5H_2O$ solution

As shown in figure 2, the average size of particles was from 20nm to 30nm when concentration of sodium thiosulfate solution was changed from 0.01mol/L to 0.03mol/L and it was rapidly increased up to 70nm over 0.04mol/L.

Under the condition in which the concentration of sodium thiosulphate solution was low, a number of nanosulfur produced per an hour was low and so the aggregation was less due to interaction but the number was high and so aggregation was occurred if the concentration of sodium thiosulfate solution was high.

Thus the proper concentration of sodium thiosulfate solution for reaction is ranged from 0.01mol/L to 0.03mol/L.

Figure 3 shows the effect of concentration of sulfuric acid solution on the average diameter.

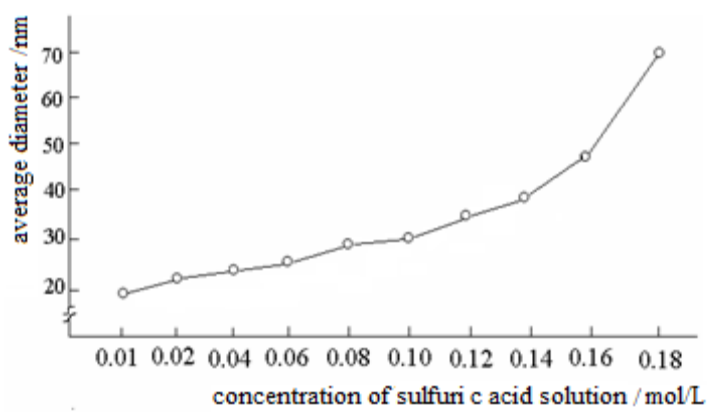

Figure 3. average diameter change of nanosulfur depending on the concentration of sulfuric acid solution

As shown in figure 3, when concentration of sulfuric acid was changed fro 0.01mol/L to 0.10mol/L, the average size was from 20nm to 30nm and it was rapidly increased more than 0.12mol/L so that it was up to 60nm at 0.18mol/L.

It can be explained similar to concentration effect of sodium thiosulfate solution on the average diameter of nanosulfur.

From this, the proper concentration of sulfuric acid solution for reaction is ranged from 0.01mol/L to 0.10mol/L.

Table 1 shows the influence of dropping & stirring rates on the average diameter of nanosulfur.

Table 1. average diameter of nanosulfur depending on the dropping & stirring velocity(nm)

| № | Dropping rate/ (drops·min$^{-1}$) | Stirring rate/(r·min$^{-1}$) | | | | | |
|---|---|---|---|---|---|---|---|
| | | 100 | 250 | 500 | 1000 | 1250 | 1500 |
| 1 | 20 | 78 | 64 | 40 | 20 | 20 | 20 |
| 2 | 40 | 89 | 73 | 50 | 25 | 27 | 25 |
| 3 | 60 | 96 | 84 | 65 | 30 | 30 | 31 |
| 4 | 80 | 105 | 92 | 70 | 45 | 46 | 46 |
| 5 | 100 | 120 | 110 | 97 | 58 | 58 | 57 |

As can be seen from table 1, the less the amount of sulfuric acid solution dropped is and the faster the stirring rate is, the smaller the average size of nanosulfur particles.

Thus the reasonable dropping rate is 20~60 drops per minute and the reasonable stirring rate is 1000rpm.

We considered the effect of reaction time on the yield of nanosulfur obtained under the same condition as described above.

Figure 4 shows the yield of nanosulfur becomes high when the reaction time is increased, but there is no chang in yield over 3 hours.

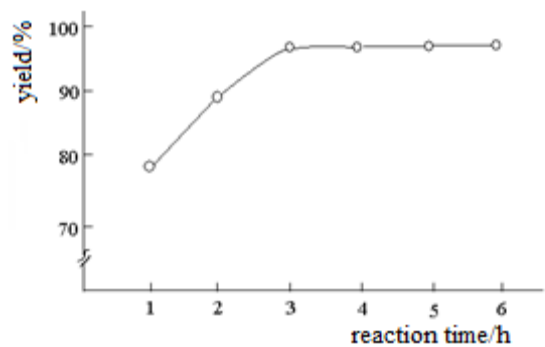

Figure 4. yield change of nanosulfur depending on the reaction time

So we can know that the proper reaction time for synthesis reaction of nanosulfur is 3 hours.

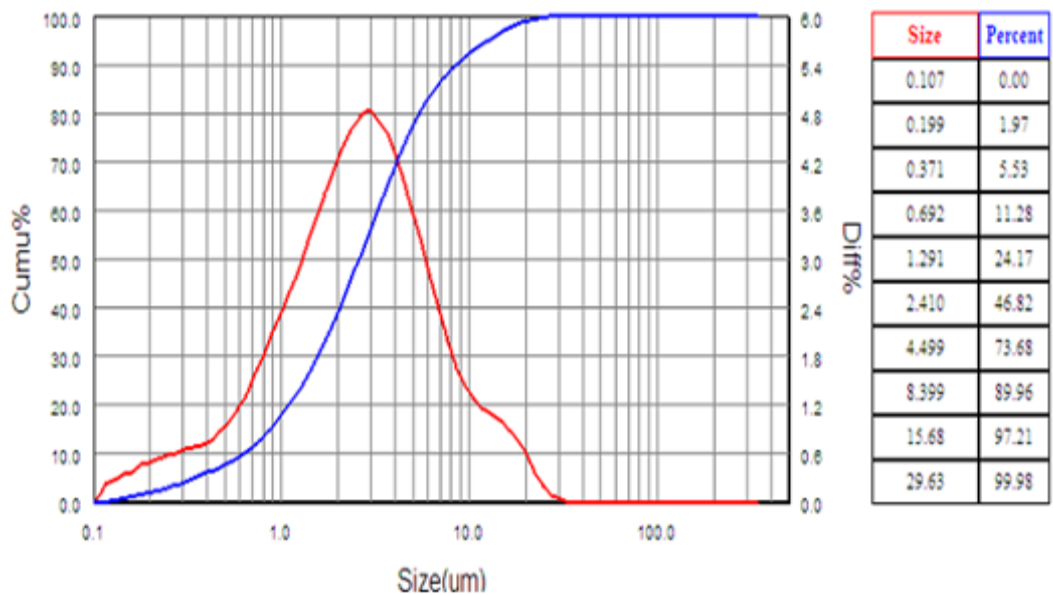

Figure 5. Cumulative curve depending on the particle size of nanosulfur-bentonite complex

As shown in figure 5, particle size of nanosulfur-bentonite complex is a range from 0.1 $\mu m$ to 20 $\mu m$.

The surface morphology of bentonite was observed under the acceralating voltage of 15kV and the magnification of X 20 000, figure 6, a).

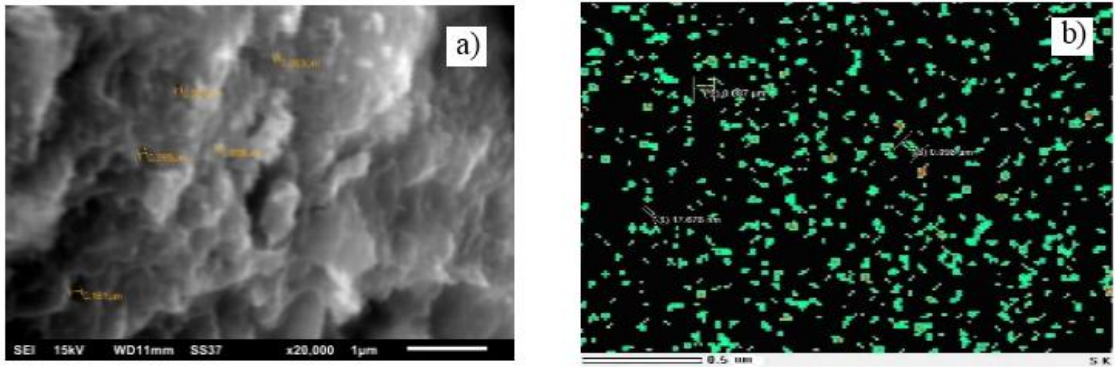

Figure 6. a) Surface morphology of nanosulfure-bentonite complex and b) Distribution graph of nanosulfur

This figure shows that the flexion of surface is remarkable and some square pores of 81nm~160nm are existed.

As a result, nanosulfur can be immersed here well.

Figure 4 a) is a distribution image of sulfur observed under the acceralating voltage of 15kV, the magnification of X 10 000 and 1024×768 pixels.

And the results show that the sulfur particles are distributed uniformly on the surface of bentonite and it does not cover completely on the surface of bentonite and nanosulfur particles are immersed in a certain distance.

## 4. Conclusion

We established the manufactured condition of nanosulfur-bentonite complexwhich nanosulfur in solution state is immersed into bentonite.

The reasonable manufactured condition of nanosulfur-bentonite complex is the reaction temperature of 10℃, the concentration of sodium thiosulfate solution 0f 0.03mol/L, the concentration of sulfuric acid solution of 0.01mol/L, the dropping rate of 20 drops per minute, the stirring rate of 1000rpm and the addition of bentonite of 30g.

In nanosulfur-bentonite complex, nanosulfur particles does not cover completely on the surface of bentonite with a certain distance and then small particle size is a range from 20nm to 30nm and the sulfur particles interfered with them is a range from 60nm to 100nm.